\documentstyle [12pt] {article}
\title{CAN COSMOLOGICAL CONSTANT BE A FORBIDDEN ZONE (GAP)
 IN QUANTUM VACUUM}

\author{Vladan Pankovi\'c$^{\ast,\sharp}$, Rade Glavatovi\'c$^\diamond$,
Simo Ciganovi\'c$^\sharp$\\
$^\ast$Department of Physics, Faculty of Sciences, 21000 Novi
Sad,\\ Trg Dositeja Obradovi\'ca 4. , Serbia, vdpan@neobee.net \\
$^\sharp$Gimnazija, 22320 Indjija, Trg Slobode 2a, Serbia\\
$^\diamond$ Military-Medical Academy, 11000 Belgrade, Crnotravska
17., Serbia \\}

\date {}
\begin{document}
\maketitle

 PACS number : 98.80.-k, 98.80.Qc

\begin {abstract}
In this work we suggest, without detailed mathematical analysis, a
hypothesis on the physical meaning of cosmological constant. It is
primarily based on a conceptual analogy with energy
characteristics of the crystal lattice structure, i.e. energy
zones theory in solid state physics. Namely, according to some
theories (holographic principle, emergent gravity etc.) it is
supposed that empty space, i.e. quantum vacuum holds a structure
like to crystal lattice. It implies a possibility of the existence
of totally occupied zones consisting of many levels of the
negative energies as well as at least one negative energy
forbidden zone, i.e. negative energy gap without any (occupied or
empty) level of the negative energy. We suppose that given
negative energy forbidden zone in the quantum vacuum represents
effectively a positive energy zone without quantum particles that
corresponds to cosmological constant. Also we suggest some other
(less extravagant) model of the cosmological constant. Here
cosmological constant is usually considered as the effect of the
quantum vacuum fluctuations where problem of the cut-off can be
solved quite simply since here integration over unlimited domain
of the quasi-momentums must be changed by integration over one,
finite "Brillouin zone".
\end {abstract}

\vspace{1cm}

It seems that recent astronomical data on the universe
acceleration are in the best agreement with theoretical
predictions on the cosmological constant that can be considered as
energy of the (quantum) vacuum [1], [2]. Nevertheless physical
meaning of the non-gravitational (anti-gravitational) vacuum
energy is quite mysterious [3]-[6]. This "Mystery of the
Millennium" [3] implies "Crying Need for Some Bright Ideas"[4].

In this work we shall suggest, without detailed theoretical
formalism, a hypothesis on the physical meaning of cosmological
constant. It is primarily based on a conceptual analogy with
energy characteristics of the crystal lattice structure, i.e.
energy zones theory in solid state physics. Namely, as it is
well-known, in some theories of the quantum gravitation, e.g. in
theories of the holographic principle [7]-[11] or emergent gravity
[12], etc., there is supposition that (gravitational) vacuum has a
space structure similar to a three-dimensional spin lattice with
distance between sides proportional to Planck length. "Let us
temporarily suppose the world is a 3 dimensional lattice of spin
like degrees of freedom. For definiteness assume the lattice
spacing is the Planck length lP and that each site is occupied
with a spin which can be in one of two states. For example a
lattice fermion field theory would be of this type." [8]

Given supposition we shall consider and further extend,
conceptually but without detailed mathematical analysis, in this
work. We shall suppose that empty space, precisely (gravitational)
vacuum has really a lattice structure like a solid state crystal
lattice. Also, we shall suppose that given (gravitational) vacuum
lattice holds a dynamics similar, at least conceptually, to the
quantum dynamics of a solid state crystal structure. It can imply
a possibility of the formation of totally occupied zones
consisting of many discrete levels with negative energies as well
as at least one negative energy forbidden zone, i.e. negative
energy gap without any (occupied or empty) level of the negative
energy. Finally, we shall suppose that given negative energy
forbidden zone in quantum vacuum represents effectively a positive
energy zone without quantum particles that corresponds to
cosmological constant. Also we shall suggest some other (less
extravagant) model of the cosmological constant. In this
alternative model cosmological constant is usually considered as
the effect of the quantum vacuum fluctuations according to vacuum
lattice structure. Here problem of the cut-off can be solved quite
simply since integration over unlimited domain of the
quasi-momentums must be changed by integration over one, finite
"Brillouin zone".

Thus suppose that empty space, i.e. (gravitational) quantum vacuum
has really a lattice structure like a solid state crystal lattice.
Further, suppose that given vacuum lattice holds "unmovable"
elements, individually denoted {\it A}, in lattice sides, that,
practically, build given lattice, like to atoms in the solid state
crystal lattice. Also, suppose that vacuum lattice holds
"movable", at least formally, elements, individually denoted {\it
a}, that represent fermions and that can move, at least formally,
through given lattice like to electrons in the solid state
crystals.

Suppose now that given vacuum lattice holds a dynamics similar, at
least conceptually, to the quantum dynamics of a solid state
crystal structure.

Suppose that in the first approximation given dynamics would be
roughly presented as "motion" (at least in the formal sense) of
{\it a}-s in a potential hole determined by {\it A}-s, i.e. by
vacuum crystal lattice. Periodical potential of given vacuum
lattice can cause that given potential hole can hold only discrete
levels of the negative energy any of which, according to
remarkable Dirac interpretation of the quantum vacuum, is
completely occupied by two {\it a}-s representing fermions. For
this reason there is no any "motion" of {\it a}-s through vacuum
lattice. All this conceptually corresponds to approximation of the
Fermi gas of the motion of electrons in the metals at 0 K
temperature when all energy levels under Fermi level are
completely occupied and when motion of the electrons is
impossible.

We shall suppose that, in the more subtle description of the
quantum vacuum lattice dynamics, dynamical interaction between
{\it A}-s, i.e. lattice and {\it a}-s, induces splitting of the
{\it a}-s negative energy levels and their grouping in the
negative energy zones. Suppose, for reason of simplicity, that
there are only two completely occupied, allowable negative energy
zones, $W_{1}$ and $W_{2}$.

Suppose that give two allowable negative energy zones are
discretely separated by one (for reason of simplicity we shall
suppose that it is unique) negative energy gap, i.e. forbidden
negative energy zone or zone without any level of negative energy
$W_{g}$ .

All this is, of course, conceptually analogous to the zone theory
in the solid state physics.

Completely occupied, allowable negative energy zones $W_{1}$  and
$W_{2}$ can be considered as usual parts of the quantum vacuum
without any quantum "motion", i.e. transition from one in the
other level within any of given zones. It is possible, of course,
suppose that some external dynamical interaction realizes
transition of some {\it a} from $W_{1}$ or $W_{2}$ in the levels
with positive energy so that, according to Dirac anti-matter
interpretation, absence, i.e. hole of {\it a} in $W_{1}$ or
$W_{2}$, can be effectively considered as an anti-particle with
positive mass-energy. Quantum motion of the hole of {\it a}
through $W_{1}$ or $W_{2}$ corresponds conceptually to quantum
motion of the anti-particle through positive energy levels.

It is quite reasonable to be supposed that positive energy in form
of quantum particle or anti-particle (including field quants)
satisfies equivalence principle and gravitates.

Forbidden negative energy zone or zone without levels with
negative energy, $W_{g}$, can be considered as an unusual part of
the (gravitational) quantum vacuum. It needs an additional
interpretation. Quantum system {\it a} from $W_{1}$ or $W_{2}$ or
from any positive energy level cannot arise in $W_{g}$. Also,
since $W_{g}$ has not any (occupied or empty) energy level, there
is no any dynamical interaction that can realize transition of
{\it a} from $W_{g}$ in $W_{1}$ or in $W_{2}$or in any positive
energy level. It, roughly speaking, means that $W_{g}$ has not any
correct presentation by individual quantum particle or
anti-particle. Nevertheless, in full spirit of the Dirac
anti-matter interpretation, absence of the negative energy levels
in $W_{g}$ can be effectively considered as the presence of some
kind of the positive energy.

It is quite reasonable to be supposed that positive energy without
quantum particle or anti-particle corresponding to $W_{g}$ acts
anti-gravitationally. Indeed, roughly speaking, forbidden zone
$W_{g}$ forbids gravitational falling of {\it a} and other quantum
particles or anti-particles in given domain of the negative
energies.

Crystal lattice structure of the (gravitational) quantum vacuum
can imply and some other (less extravagant) possibility for
cosmological constant modeling. Namely, it can be again supposed
that given vacuum lattice structure and corresponding periodical
potential energy determine a translator symmetric dynamics like to
Bloch-Brillouin dynamics in the quantum solid state physics. Then
physical states can have form of the quasi-free waves
corresponding to quantum fluctuations of the vacuum. Given waves ,
i.e. fluctuations can hold quasi-momentums (different, exactly
speaking from usual momentums). Energy spectrum obtains then a
zone structure since energy of the states is no more a completely
continuous function of the quasi-momentums. Namely, then there are
different allowable energy zones (any of which represents a
continuous function of the quasi-momentums from corresponding
finite intervals or "Brillouin zones") discretely separated by
forbidden energy zones. In this case problem of the cut-off by
defining of the cosmological constant by vacuum fluctuations [5]
can be solved quite simply since here integration over unlimited
domain of the quasi-momentums must be changed by integration over
one, finite "Brillouin zone".

Finally, we can shortly conclude and repeat the following. In this
work we suggested, without detailed mathematical analysis, a
hypothesis on the physical meaning of cosmological constant. It is
supposed that empty space, i.e. quantum vacuum holds a structure
like to crystal lattice. It implies a possibility of the existence
of totally occupied zones consisting of many levels of the
negative energies as well as at least one negative energy
forbidden zone, i.e. negative energy gap without any (occupied or
empty) level of the negative energy. We supposed that given
negative energy forbidden zone represents effectively a positive
energy zone without quantum particles that corresponds to
cosmological constant. Also we suggested an alternative model of
the cosmological constant. In this model cosmological constant is
usually considered as the effect of the quantum vacuum
fluctuations according to vacuum lattice structure. Here problem
of the cut-off can be solved quite simply since integration over
unlimited domain of the quasi-momentums must be changed by
integration over one, finite "Brillouin zone".

\vspace {1.5cm}

This work is dedicated to memory of Mica Mali.

\section {References}

\begin {itemize}

\item [[1]] D. N. Spergel et al., Astrophys. J. Supp. {\bf 146} (2003) 175 ; astro-ph/0302209
\item [[2]] D. N. Spergel et al., Astrophys. J. Supp. {\bf 170} (2007) 337 ; astro-ph/0603449
\item [[3]] T. Padmanabhan, {\it Dark Energy: Mystery of the Millennium}, astro-ph/0603114 and references therein
\item [[4]] T. Padmanabhan, {\it Darker Side of the Universe …  and the Crying Need for Some Bright Ideas}, astro-ph/0510492 and references therein
\item [[5]] V. Sahni, {\it Dark Matter and Dark Energy}, astro-ph/0403324
\item [[6]] R. Durrer, R. Maartens, {\it Dark Energy and Dark Gravity: Theory Overview}, astro-ph/0711.0077 and references therein
\item [[7]] G. t'Hooft, {\it Dimensional Reduction in Quantum Gravity},gr-qc/9310026
\item [[8]] L. Susskind, {\it World as a Hologram}, hep-th/9409089
\item [[9]] W. Fishler, L. Susskind, {\it Holography and Cosmology}hep-th/9806039
\item [[10]] R. Bousso, {\it The holographic principle}, hep-th/0203101 and references therein
\item [[11]] S. de H. Olle, {\it Quantum Gravity and Holographic Principle}, hep-th/0107032
\item [[12]] H. S. Yang, {\it Emergent Gravity and the Cosmological constant Problem}, hep-th/0711.2797

\end {itemize}

\end {document}